\documentclass[twoside]{article}
\usepackage{medias}
\ifpdf
   \usepackage{graphicx}
\else
   \usepackage{graphicx}
\fi

\usepackage{bigstrut}
 \usepackage{alltt}

\begin{document}
\English

\title{Exploring mutexes, the Oracle\textregistered   RDBMS retrial spinlocks}
\author{Nikolaev~A.\,S.}
\organization{RDTEX LTD, Protvino, Russia 

\vspace*{5mm}
{\large Proceedings of International Conference on Informatics MEDIAS2012. 

Cyprus, Limassol, May 7--14, 2012. ISBN 978-5-88835-023-2}}
\email{Andrey.Nikolaev@rdtex.ru, {http://andreynikolaev.wordpress.com}}

\abstract{
Spinlocks are widely used in database engines for  processes synchronization.
KGX mutexes is new retrial spinlocks appeared  in contemporary Oracle\textregistered\ versions for submicrosecond synchronization.
The mutex contention is frequently observed in highly concurrent OLTP environments.

This work explores how Oracle mutexes operate, spin, and sleep. It develops predictive mathematical model and discusses parameters and statistics related to mutex performance tuning, as well as results of contention experiments.
}
\titleRus{Исследование мьютексов, спинблокировок с повторными вызовами  в СУБД Oracle\textregistered\ }
\authorRus{Николаев~А.\,С.}
\organizationRus{Протвино, ЗАО РДТЕХ}
\abstractRus{Спинблокировки широко используются в  системах управления базами данных для синхронизации процессов.
В современных версиях СУБД Oracle появились KGX мьютексы -  новый тип  спинблокировок  с повторными вызовами ориентированный на синхронизацию в субмикросекундных масштабах. Конкуренция за мьютексы часто наблюдается в высоконагруженных OLTP средах.

В работе  исследуются особенности циклирований и ожиданий мьютексов. На основе построенной математической модели обсуждаются параметры и статистики связанные с настройкой производительности мьютексов, а также  результаты экспериментов.}
\maketitle

\section{I. Introduction}
\newcommand{\Ffactor}{{\mathsf{F}_c}}
\newcommand{\hpdf}{p}
\newcommand{\muf}{}

According to  Oracle\textregistered\   documentation \cite{Concepts112}  mutex is:
''{\em A mutual exclusion object \ldots that
prevents an object in memory from aging out or from being corrupted \ldots}''.

Huge  Oracle  RDBMS  instance  contains  thousands  processes
accessing the shared memory. This shared memory  named ''System Global Area'' (SGA)
consists of millions cache, metadata and results structures.
Simultaneous access to these structures is synchronized by Locks, Latches and KGX Mutexes:
\begin{figure}[H]
\begin{center}
\includegraphics[width=\linewidth]{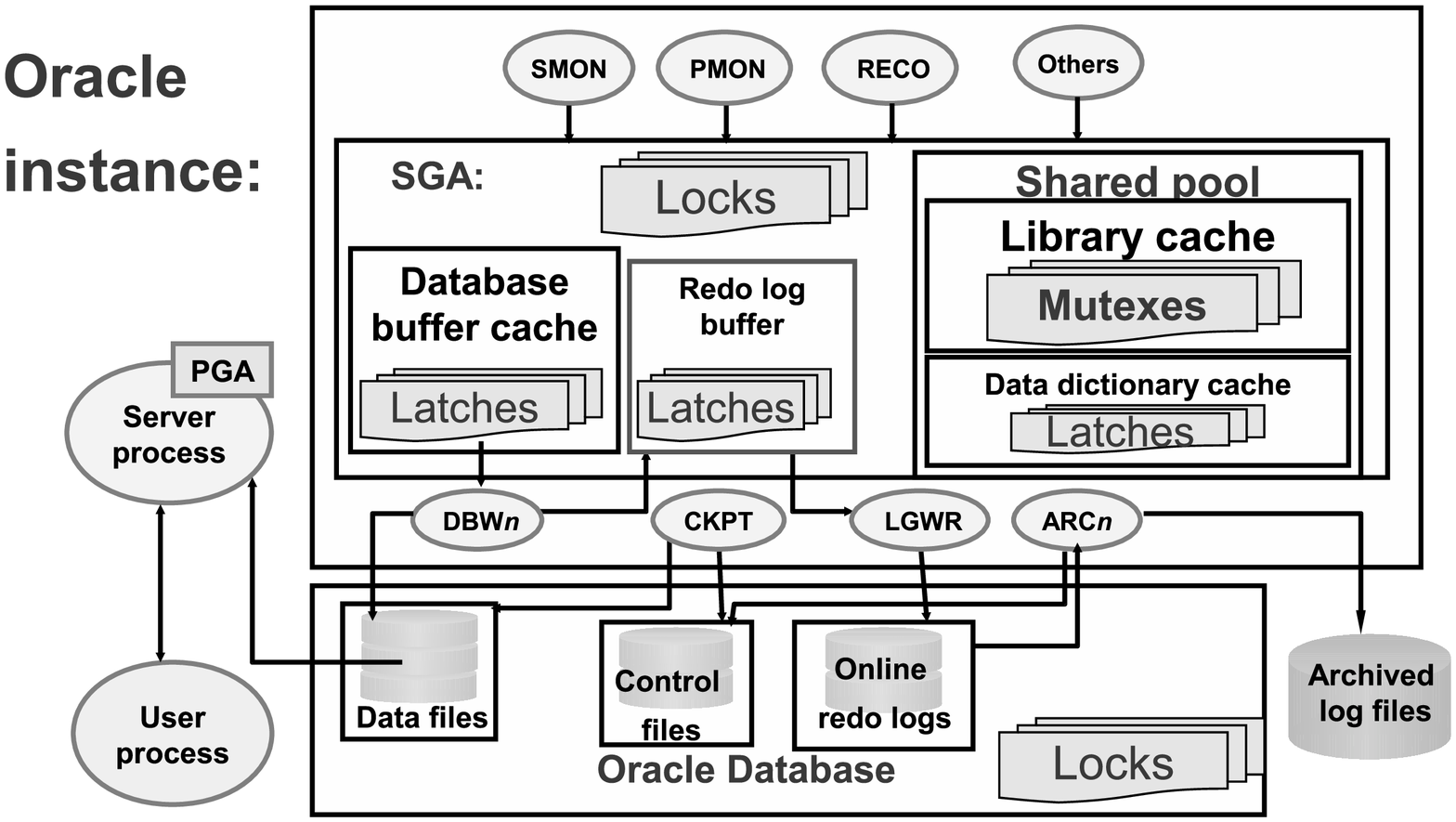}
\vspace*{-5mm}
\caption{Oracle\textregistered\  RDBMS architecture}\label{fig:arc}
\end{center}
\end{figure}

\vspace*{-5mm}
 Latches and  mutexes are the Oracle   realizations of   spinlock concept.
My previous article \cite{latches2011} explored latches, the traditional Oracle spinlocks known since  1980th.
Process requesting the latch spins 20000 cycles polling its location.
If unsuccessful, process joins a queue and uses wait-posting to be awaken  on latch release.

The goal of this work is to explore  the  newest Oracle spinlocks  --- mutexes.
The mutexes were introduced in 2006 for synchronization inside Oracle Library Cache.
Table \ref{tab:scomp} compares Oracle internal synchronization mechanisms.

\textbf{Wikipedia} defines the spinlock as {\em ''\ldots a lock where the thread simply waits in a loop (''spins'') repeatedly
checking until the lock becomes available. As the thread remains active but isn't performing a useful task,
the use of such a lock is a kind of busy waiting''}.

Use of spinlocks for multiprocessor synchronization was first introduced by Edsger Dijkstra in \cite{Dijkstra65}.
 Since that time, the mutual exclusion algorithms were significantly advanced.
Various sophisticated spinlock realizations (TS, TTS, Delay, MCS, Anderson, etc.) were proposed and evaluated.
The contemporary review of these algorithms may be found in \cite{Anderson2003}

Two general spinlock types exist:
\begin{description}
\item[System] spinlocks that protect critical OS structures. The kernel  thread cannot wait  or yield the CPU. It must loop until success.
Most mathematical models explore this spinlock type. Major metrics to optimize system spinlocks
are frequency of atomic operations (or Remote Memory References)  and shared bus utilization.
\item[User] application spinlocks  like Oracle latches and mutexes that protect user level structures.
It is more efficient to poll the mutex for several  microseconds rather than pre-empt the thread doing 1 millisecond context switch.
The metrics to optimize are spinlock acquisition CPU and elapsed times.
\end{description}

In Oracle versions 10.2 to 11.2.0.1 (or, equivalently, from 2006 to 2010) the mutex spun using atomic {\em{}''test-and-set''} operations.
According to Anderson classification \cite{Anderson90} it was a \textbf{TS} spinlock.
Such spinlocks may induce the Shared Bus saturation and affect performance of other memory operations.

Since version 11.2.0.2 processes poll the mutex location  nonatomically and only use
 atomic instructions to finally acquire it.  The contemporary mutex became a \textbf{TTS} ({\em{}''test-and-test-and-set''}) spinlock.

  System spinlocks  frequently use more complex structures than TTS.  Such algorithms, like famous MCS spinlocks \cite{MCS}
were optimized for ~100\% utilization.
For the current state of system spinlock theory see \cite{Art08}.

If user spinlocks are holding for a long time, for example due to OS preemption, pure spinning becomes ineffective and wastes CPU.
To overcome this, after 255 spin cycles the mutex sleeps  yielding the processor to other workloads and then retries. The sleep timeouts are determined by mutex {\em wait scheme}.

 Such {\em{} spin-sleeping} was first introduced in \cite{oust82} to achieve balance between CPU time lost by spinning and context switch overhead.

From the queuing theory point of view such systems with repeated attempts are retrial
queues. More precisely, in the retrial system the request that finds the server busy upon arrival leaves the service area and joins a retrial group ({\em orbit}). After some  time this
request  will have a chance to try its luck again. There exists an extensive literature on the
retrial queues. See \cite{falin97,artalejo08} and references therein.

The mutex retrial spin-sleeping algorithm significantly differs from the
FIFO spin-blocking used by Oracle latches \cite{latches2011}.
The spin-blocking was  explored in  \cite{agarwal91,karlin90,boguslavsky91}.
Its robustness in contemporary environments  was recently investigated in \cite{look2009}.

Historically the mutex contention issues were hard to diagnose and resolve \cite{lp}.
The mutexes are much less documented then needed and evolve rapidly.
Support engineers definitely need more mainstream science support to predict the results of changing mutex parameters.
 This paper summarizes author's work on the subject. Additional details may be found in my blog \cite{asn}.


\begin{table}[t]
\centering
\small{\setlength{\extrarowheight}{1.5pt}
\begin{tabular}{|p{0.25\linewidth}|p{0.15\linewidth}|p{0.20\linewidth}|p{0.21\linewidth}|}
\hline
&\textbf{Locks}&\textbf{Latches}&\textbf{Mutexes}\\
\hline
\textbf{Access}&Several \- Modes&Types and Modes&\textbf{Operations}\\
\hline\textbf{Acquisition}&FIFO&SIRO spin FIFO wait&\textbf{SIRO} \\
\hline\textbf{SMP Atomicity}&No&Yes&\textbf{Yes}\\
\hline\textbf{Timescale}&Milli\-seconds&Micro\-seconds&{\textbf{Sub}Micro\-seconds}\\
\hline\textbf{Life cycle}&Dynamic&Static&\textbf{Dynamic}\\
\hline
\end{tabular}
}\normalsize
\caption{Serialization mechanisms in Oracle}
\label{tab:scomp}
\end{table}

\section{II. Oracle\textregistered\  RDBMS  Performance Tuning overview}

Before discussing the mutexes, we need some introduction.
During the last 33 years, Oracle evolved from the first  one-user SQL database to the most advanced contemporary RDBMS engine.
Each version introduced  performance and concurrency advances:
{\small
\begin{description}
\item[v. 2] (1979): the first commercial SQL RDBMS.
\item[v. 3] (1983): the first database to support SMP.
\item[v. 4] (1984): read-consistency, Database Buffer Cache.
\item[v. 5] (1986): Client-Server, Clustering, Distributed Database, SGA.
\item [v. 6] (1988): procedural language (PL/SQL), undo/redo, latches.
\item [v. 7] (1992): Library Cache, Shared SQL, Stored procedures, 64bit.
\item [v. 8/8i] (1999): Object types, Java, XML.
\item [v. 9i] (2000): Dynamic SGA, Real Application Clusters.
\item [v. 10g] (2003): Enterprise Grid Computing, Self-Tuning, \textbf{mutexes}.
\item [v. 11g] (2008): Results Cache, SQL Plan Management, Exadata.
\item [v. 11gR2] (2011): \ldots \textbf{Mutex wait schemes. Hot object copies}.
\item [v. 12c] (2012): {\em{} Cloud}\ldots
\end{description}
}

As of now, Oracle  is the most  widely used SQL RDBMS. In majority of workloads  it  works  perfectly.
However, quick search finds more then 100 books devoted to Oracle performance tuning on
 Amazon \cite{lewis,adams,millsap}. Dozens conferences covered this topic every year. Why Oracle needs such a tuning?

The main reason is complex and variable workloads. Oracle is working in very different environments
ranging from huge OLTPs, petabyte OLAPs to hundreds multi-tenant databases running on one server.
Every high-end database is unique.

For the ability to work in such diverse conditions  Oracle RDBMS has  complex internals. To get the most out of hardware  we need precise tuning.
Working at Support, I cannot underestimate the importance of developers and database administrators education in this field.

In order to diagnose performance problems Oracle instrumented its software.
Every Oracle session keeps many statistics counters describing {\em{}''what was done''}.
 Oracle Wait Interface (OWI) \cite{millsap} events describe
 {\em ''why the session waits''} and complement the statistics.

Statistics, OWI and  data from internal (''{\em fixed}'') \textbf{X\$} tables are used by
Oracle diagnostics and visualization tools.

This is the traditional framework of Oracle performance tuning. However, it was not effective enough in spinlocks troubleshooting.

\subsection{The DTrace.}

To observe the mutexes work and short duration events, we need something like stroboscope in physics. 
Likely, such tool exists in Oracle Solaris\texttrademark. This is DTrace, Solaris 10 Dynamic Tracing framework \cite{dtrace}.

DTrace is event-driven, kernel-based instrumentation that can
see and measure all OS activity. It defines  \textbf{probes}
 to trap and handlers (\textbf{actions}) using dynamically interpreted C-like language.
No application changes needed to use DTrace.  This is very similar to triggers in database technologies.

DTrace can:
\begin{itemize}
\item{} Catch any event inside Solaris and function call inside Oracle.
\item Read and change any address location in-flight.
\item Count the mutex spins, trace the mutex waits, perform experiments.
\item Measure times and distributions up to microsecond precision.
\end{itemize}

 Unlike standard tracing tools, DTrace works in Solaris kernel.
When process entered probe function, the execution went to Solaris kernel and the
DTrace filled buffers with the data.
Kernel based tracing is more stable and have less overhead then userland.
DTrace sees all the system activity and can  account the
 time associated with kernel calls, scheduling, etc.

In the following sections describing Oracle performance tuning are interleaved by mathematical estimations.

\section{III. Mutex spin model}

\begin{figure}[t]
\begin{center}
\includegraphics[width=\linewidth]{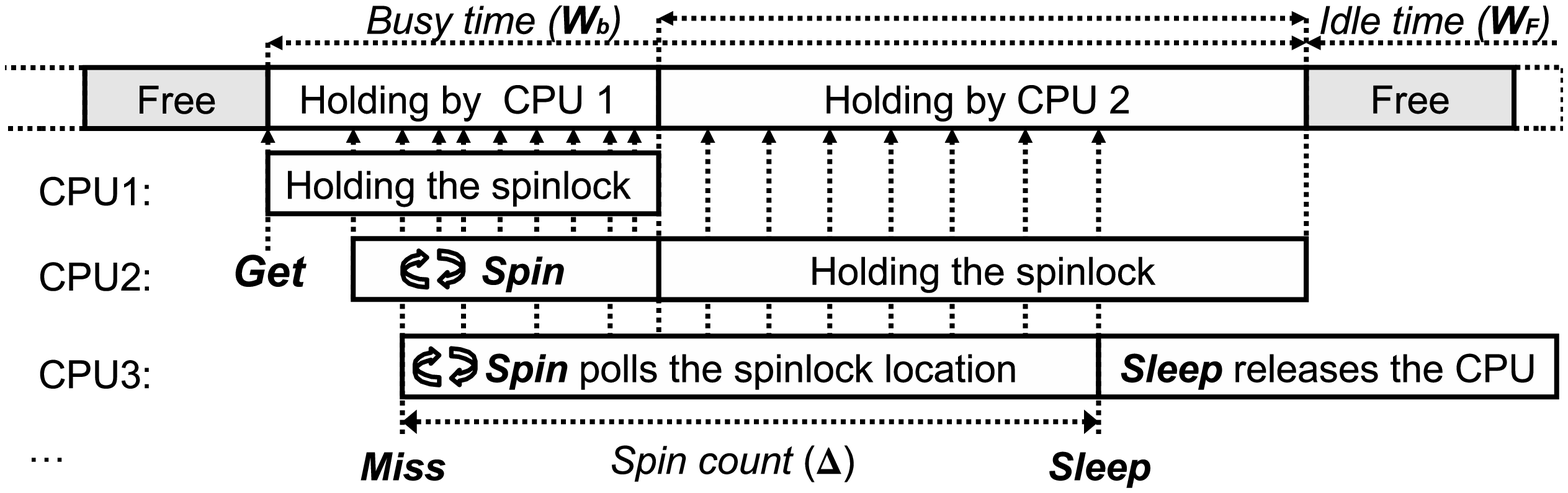}
\end{center}
\vspace*{-5mm}
\caption{Oracle mutex workflow.}\label{fig:mwflow}
\end{figure}

The Oracle mutex workflow schematically visualised in fig. \ref{fig:mwflow}.
 The Oracle process:
\begin{itemize}
\item Uses atomic hardware instruction for mutex \emph{Get}.
\item If \emph{missed}, process spins by polling mutex location during \emph{spin get}.
\item Number of spin cycles is bounded by \emph{spin count}.
\item If spin get not succeeded, the process acquiring mutex \emph{sleeps}.
\item During the sleep the process may wait for already free mutex.
\end{itemize}
Oracle counts \emph{Gets} and \emph{Sleeps} and we can measure \emph{Utilization}.

This section introduces the mathematical model used to forecast  mutex behaviour. It extends the model used in
\cite{agarwal91,latches2011} 
 for general holding time distribution  and  TTS spinlock concurrency.

Consider a general stream of mutex holding events. The mutex memory location have been  changed by  sessions  at time $H_k$, $k\in \mathcal{N}$ using atomic instruction.
This  instruction blocked the shared bus and succeeded only when memory location is free.

After  acquisition the session  will hold the mutex for time  $x_k$ distributed with  p.d.f. $\hpdf(t)$.
I assume that incoming stream is Poisson with rate $\lambda$ and  $H_k$ (and $x_k$) are generally independent  forming renewal process.
Furthermore, I assume here the existence of at least second moments for all the distributions.

 The mutex acquisition request  at  time $T_m$, $m\in \mathcal{N}$  succeeds immediately  if it finds the mutex  free.
Due to \emph{Serve-In-Random-Order} nature of spinlocks, there is no simple relation between $m$ and $k$.

If the mutex was busy at  time $T_m$:
\[ H_k<T_{m}<H_k+x_k \quad \textrm{for some}\; k,\]
 then \emph{miss} occurred. According to PASTA property the \emph{miss} probability is equal to mutex utilization $\rho$.

 Missing session will \emph{spin} polling the mutex location up to time $\Delta$ determined by {\bf\_mutex\_spin\_count} parameter. The    initial ''\emph{contention free}'' approximation assumes that no other requests arrive during  the spin ($\lambda \Delta \ll 1$) and  the session  acquires the mutex  if it  become free while spinning.
Therefore  the spin succeeds if:
\begin{displaymath}
T_m+\Delta > H_k + x_k.
\end{displaymath}
If the mutex was not released during $\Delta$, the session  sleeps.

According to classic considerations of  renewal theory \cite{kleinrock,Cox},
 incoming requests   peek up the holding   intervals with p.d.f.:
 \begin{equation} p_h=\frac{1}{S} x \hpdf(x),\label{eq:phx}
\end{equation}
and   observes the transformed mutex holding time distribution.  Here $S=\Expect(x)$ is the average mutex holding time. The p.d.f. and average of residual holding time  is well-known:
 \begin{equation}
\begin{array}{l}
p_r(x)= \frac{1}{S} \int\limits_x^\infty p(t) \,\mathrm{d}t\\
S_r=\frac{1}{2S } \Expect( t^2 )\label{eq:pr}
\end{array}
\end{equation}

The spin time distribution (conditioned on miss) follows the c.d.f. $\Prob_r$, but has  a discontinuity \cite{latches2011} at $t= \Delta$   because the session acquiring latch  never  spins more than $ \Delta$. The magnitude of this discontinuity is  the overall probability that residual mutex holding time will be greater then $\Delta$. Corresponding p.d.f. is:
\begin{equation} p_{sg}(x)=\hpdf_r(x) H(\Delta-x) + \mathsf{Q}_r(\Delta)  \delta(x-\Delta)
\end{equation}
Here $ H(x)$ and $ \delta(x)$ is Heaviside step and bump functions correspondingly.
\vspace{5mm}

\paragraph{The spinlock observables}

Oracle statistics allows measuring of   \emph{spin inefficiency} (or \emph{sleep ratio})  coefficient $k$. This is the probability do not acquire mutex during the spin.  Another crucial quantity is $\Gamma$ - the average CPU time spent while spinning for the mutex:
\begin{equation}
\left\{\begin{array}{l}
k_0 = Q_r(\Delta) = \int_\Delta^\infty p_r(t)  \,\mathrm{d}t = \frac{1}{S}\int\limits_\Delta^\infty\ Q(t) \mathrm{d}t \\
 \Gamma=\int\limits_0^\infty t p_{sg}(t) \; \mathrm{d}t = \frac{1}{S}\int\limits^\Delta_0 \mathrm{d}t \int\limits^\infty_t Q(z) \; \mathrm{d}z\\
\end{array}\right.
\end{equation}
Here subscript 0 denotes "contention free" approximation.
Using \eqref{eq:pr} for distributions with finite dispersion this expressions can be rewritten in two ways \cite{latches2011}.

\paragraph{Low spin efficiency region}

The first form  is suitable for  the region of low spin efficiency   $\Delta \ll S$:
\begin{equation}
 \left\{\begin{array}{l}
k_0= 1- \frac{\Delta}{S} + \frac{1}{S} \int\limits^\Delta_0 (\Delta - t )  p(t) \; \mathrm{d}t\\
\Gamma=  \Delta - \frac{\Delta^2 }{2 S} + \frac{1}{2S} \int\limits^\Delta_0 (\Delta - t)^2  p(t) \; \mathrm{d}t  \\
\end{array}
\right.
\end{equation}

From the above expressions it is clear that  spin probes the mutex holding time distribution around the origin.

Other parts of mutex holding time p.d.p. impact spin efficiency and CPU consumption only through the average holding time $S$.
This allows to estimate how these quantities depend upon  \textbf{\_mutex\_spin\_count} (or $\Delta$) change.
 If process never releases mutex immediately ($p(0)=0$) then
\[
 \left\{\begin{array}{l}
k= 1- \frac{\Delta}{S} + O(\Delta^3)\\
\Gamma=  \Delta - \frac{\Delta^2 }{2S} +  O(\Delta^4)\\
\end{array}
\right.
\]

For Oracle performance tuning purpose we need to know what will happen if we double the $\Delta$:

 {\em{}In low efficiency region doubling the spin count will double the
number of efficient spins and also double the CPU consumption.}

\paragraph{High spin efficiency region}

In high efficiency region the sleep cuts off the tail of spinlock holding time
distribution:
\begin{equation}
 \left\{\begin{array}{l}
k_0= \frac{1}{S} \int\limits^\infty_\Delta (t-\Delta)  p(t) \, \mathrm{d}t\\
\Gamma=  \frac{\Expect(t^2 )}{2 S}   - \frac{1}{2S} \int\limits^\infty_\Delta (t-\Delta)^2  p(t) \, \mathrm{d}t = S_r - T_r  \\
\end{array}
\right.\label{eq:nocont}
\end{equation}
here $T_r$ is the residual after-spin holding time. This quantity  will be used later.

Oracle normally operates in this region of small  sleeps ratio.
Here the spin count is greater than number of instructions protected by mutex $ \Delta \ge S$.
The spin time is bounded by both  the "residual  holding time" and the spin count:

 \begin{displaymath}
 \Gamma < \min (\;S_r \;,\; \Delta)
\end{displaymath}
The sleep prevents process from waste CPU for spinning on heavy tail of mutex holding time distribution

\subsection{Concurrency model}
\begin{figure}[t]
\begin{center}
{\includegraphics[width=0.6\linewidth]{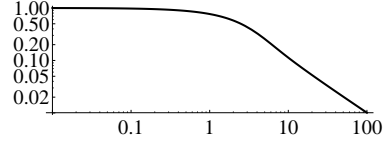}}
\end{center}
\vspace{-7mm}
\caption{The concurrency formfactor $\Ffactor(x)$.}\label{fig:Fx}
\end{figure}

In the real world several processes may  spin  on different processors concurrently. After the mutex release all these sessions will issue atomic \emph{Test-and-Set} instructions to acquire the mutex. Only one instruction succeeds. What should  the other sessions do? This is the principal question for  hybrid TTS spinlocks.

The session may either \emph{continue the spin} upto  {\bf \_mutex\_spin\_count} or it may proceed to \emph{sleep} immediately. The sleep seems reasonable because the session knows that spinlock just became busy.


For further estimations I will use the second scenario.  After the mutex release only one spinning session acquires  it according to SIRO discipline, while all other sessions sleep.
This is interesting queuing  discipline that to my knowledge has not been explored in literature. Its C pseudocode looks like:
{\bf \begin{verbatim}
while(1){ i:=0;
    while(lock<>0 && i<spin_count) i:=i+1;
    if(Test_and_Set(lock))return SUCCESS;
    Sleep();
}
\end{verbatim}}

The time just after the mutex release is the \emph{Markov regeneration point}. All the spinning behavior after this time is independent of previous history.

Consider mutex holding interval of length $x$ containing at least  one (tagged) incoming request for mutex from Poisson stream with rate $\lambda$.  The conditional probability that this interval will contain exactly $n$ incoming requests is:
\[
 \frac{1}{1-e^{-\lambda x}} \frac{(\lambda x)^n}{n!} e^{-\lambda x}, \quad n\ge1.
\]

The session will acquire   mutex at these conditions with probability $1/n$. Overall probability for tagged session to acquire mutex is:

\begin{equation}
\Ffactor(x)= \frac{1}{e^{\lambda x}-1} \sum_{n=1}^\infty  \frac{(\lambda x)^n}{n \, n!}  =
 - \frac{\mathrm{Ein}(-\lambda x)}{e^{\lambda x}-1}
\end{equation}
Here $\mathrm{Ein}(z)
= \int_0^z (1-e^{-y})\frac{\mathrm{d}y}{y} $ is the \emph{Entire  Exponential integral}  \cite{dlmf}.

The \emph{concurrency formfactor} $\Ffactor(x)$ is a smooth monotonically decreasing function with asymptotics $\Ffactor(x)=1-x/2+O(x^2)$ around 0 and $\Ffactor(x)\approx 1/x+O(1/x^2), x\to\infty$. It may be efficiently  approximated by rational functions \cite{luke69b}. Fig. \ref{fig:Fx} shows its  Log-Log plot.

 Normally mutex operates in region $\lambda x \ll 1$ and value of  formfactor $\Ffactor$ is very close to 1.

\begin{figure}[H]
\begin{center}
{\includegraphics[width=0.95\linewidth]{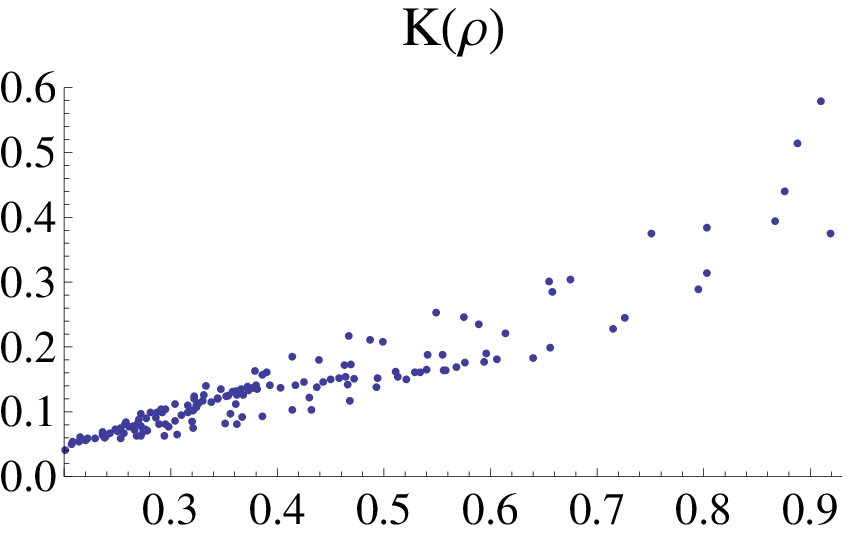}}
\end{center}
\vspace{-3mm}
\caption{$k(\rho)$.}\label{fig:krho}
\end{figure}

According to \eqref{eq:phx} the probability that missing request observe the holding interval from $x$ to $x+\mathrm{d}x$,  its residual holding time will be  from $t$ to $t+\mathrm{d}t$, $t\le x$ and it will concurrently acquire mutex on release is:
\[
\mathrm{d}\Prob = \frac{1}{S} \hpdf(x)
\> \Ffactor\!\left(\lambda  \min{(x,\Delta)} \right) \; \mathrm{d}x \, \mathrm{d}t
\]
Therefore, the \emph{spin inefficiency} or probability not to acquire mutex by spin will be:
\begin{equation}
k=1-\frac{1}{S} \int_0^\Delta \mathrm{d}t \int_t^\infty \hpdf(x)
\> \Ffactor\!\left(\lambda  \min{(x,\Delta)} \right) \;\mathrm{d}t
\end{equation}
Changing the integrations order we have:
\begin{equation}
k=1-\frac{1}{S} \int_0^\infty  \min{(x,\Delta)} \hpdf(x)
\> \Ffactor\!\left(\lambda  \min{(x,\Delta)} \right) \,\mathrm{d}x
\end{equation}
Comparing with \eqref{eq:nocont} we can outline the contention contribution:
\begin{equation}
k=k_0 +  \frac{1}{S} \int_0^\Delta \mathsf{Q}(x) \frac{\partial}{\partial x} \left( x\left(1-\Ffactor(\lambda x)\right)\right)\,\mathrm{d}x
\end{equation}
Using the formfactor asymptotics in low contention region $\lambda\Delta\ll1$ and Little law $\rho=\lambda S$ we can estimate how the spin efficiency depends on mutex utilization:
\[k=k_0+ \frac{\rho}{S^2}\int_0^\Delta x  \mathsf{Q}(x)\mathrm{d}x + o((\lambda \Delta)^2)
\]
Data of mutex contention experiments (Fig. \ref{fig:krho}) roughly agree with this linear approximation.

\section{IV. How Oracle requests the mutex}
This section discusses  Oracle mutex internals beyond the documentation.
In order to explore mutexes   we need reproducible testcases.

Each time the Oracle session executes SQL operator, it needs to \emph{pin} the cursor in library cache using mutex. "True" mutex contention arises when the same SQL operator executes concurrently at high frequency. Therefore,  simplest testcase for ''{\muf Cursor: pin S}'' contention should look like:
 \begin{alltt}{
   for i in 1..1000000 loop
     execute immediate
         'select 1 from dual where 1=2';
    end loop;}
\end{alltt}
The script uses PL/SQL loop to execute  fast SQL operator one million times.
Pure ''{\muf Cursor: pin S}''  mutex contention arises when I  execute this script by several simultaneous concurrent sessions.

It is worth to note that {\bf session\_cached\_cursors} parameter value must be nonzero to avoid soft parses. Otherwise, we will see contention for ''{\muf Library Cache}'' and ''{\muf hash table}'' mutexes also.
Indeed it is enough to  disable session cursor cache and add dozen versions of the SQL to  induce  the 
''{\muf Cursor: mutex S}'' contention.

Similarly, ''{\muf Library cache: mutex X}'' contention arises when anonymous PL/SQL block executes concurrently at high frequency.
 \begin{verbatim}
 for i in 1..1000000 loop
  execute immediate 'begin demo_proc();end;';
 end loop;
\end{verbatim}
Many other mutex contention scenarios possible. See blog \cite{asn}.

Table \ref{tab:mcomp} describes   types of mutexes in contemporary Oracle. The ''{\muf Cursor Pin}'' mutexes act as pin counters for library cache objects (e.g.
child cursors) to prevent their aging out of shared pool.
''{\muf Library cache}'' cursor and bucket mutexes protect KGL locks and static
library cache hash structures.
 The ''{\muf Cursor Parent}'' and ''{\muf hash table}'' mutexes protect parent cursors during
parsing and reloading.

\begin{table}[t]
\centering
\small{\setlength{\extrarowheight}{1.5pt}
\begin{tabular}{|p{0.15\linewidth}|p{0.3\linewidth}|p{0.40\linewidth}|}
 \hline
\textbf{Type\_id}&\textbf{Mutex\_type}&\textbf{protects}\\
\hline
\hfill\textbf{7}&{\em Cursor Pin}&Pin cursor  in memory\\
\hline\hfill\textbf{8}&hash table&Cursor management\\
\hline\hfill\textbf{6}&Cursor Parent&\ldots\\
\hline\hfill\textbf{5}&Cursor Stat&\ldots\\
\hline\hfill\textbf{4}&{\em Library Cache}&Library cache management\\
\hline\hfill\textbf{3}&HT bucket mutex (kdlwl ht)&SecureFiles management\\
\hline\hfill\textbf{2}&SHT bucket mutex&\ldots\\
\hline\hfill\textbf{1}&HT bucket mutex&\ldots\\
\hline\hfill\textbf{0}&FSO mutex&\ldots\\\hline
\end{tabular}
}
\normalsize
\caption{Mutex types in Oracle 11.2.0.3}\label{tab:mcomp}
\end{table}

The mutex address can be obtained from   {\bf x\$mutex\_sleep\_history}  Oracle table.  Such {\em ''fixed''} tables externalize internal Oracle structures to SQL. Due to dynamic nature of mutexes, Oracle does not have any fixed table like {\bf v\$latch} containing data about all mutexes.

According to Oracle documentation, this table  is circular buffer containing data about  latest mutex waits. However, my experiments demonstrated   that it is actually hash array in SGA. The hash key of this array is likely to depend on mutex address and the ID of blocking session. Row for each next sleep for the same mutex and blocking session  replaces the row for previous sleep.

\paragraph{Mutexes in memory}

We can examine mutex using {\bf oradebug peek} command. It shows memory contents:
{\small\bf\begin{verbatim}
SQL> oradebug peek 0x3F119B5A8 24
  [3F119B5A8, 3F119B5CC) =
  00000016 00000001 0000001D 000015D7 382DA701 03
    SID     refcnt     gets    sleeps   idn    op
\end{verbatim} }
According to Oracle documentation the mutex structure contains:
\begin{itemize}
\item{\em Atomically modified value} that consists of two parts:
\begin{description}
\item[Holding SID.] Top 4 bytes contain SID of session currently holding the
mutex eXclusively or modifying it. It is session number 0x16 in the above example.
\item[Reference count.] Lower 4 bytes represent the number of sessions
currently holding the mutex in Shared mode (or is in-flux).
\end{description}
\item{\em GETS} - number of times the mutex was requested
\item{\em SLEEPS} - number of times sessions slept for the mutex
\item{\em IDN} - mutex Identifier. Hash value of library cache object protected by mutex or hash bucket number.
\item{\em OP} - current mutex operation.
\end{itemize}

Oracle session changes the mutex state through  dynamic structure called Atomic Operation Log (AOL).\vspace*{-2mm}
\begin{figure}[H]
\begin{center}
\includegraphics[width=0.95\linewidth]{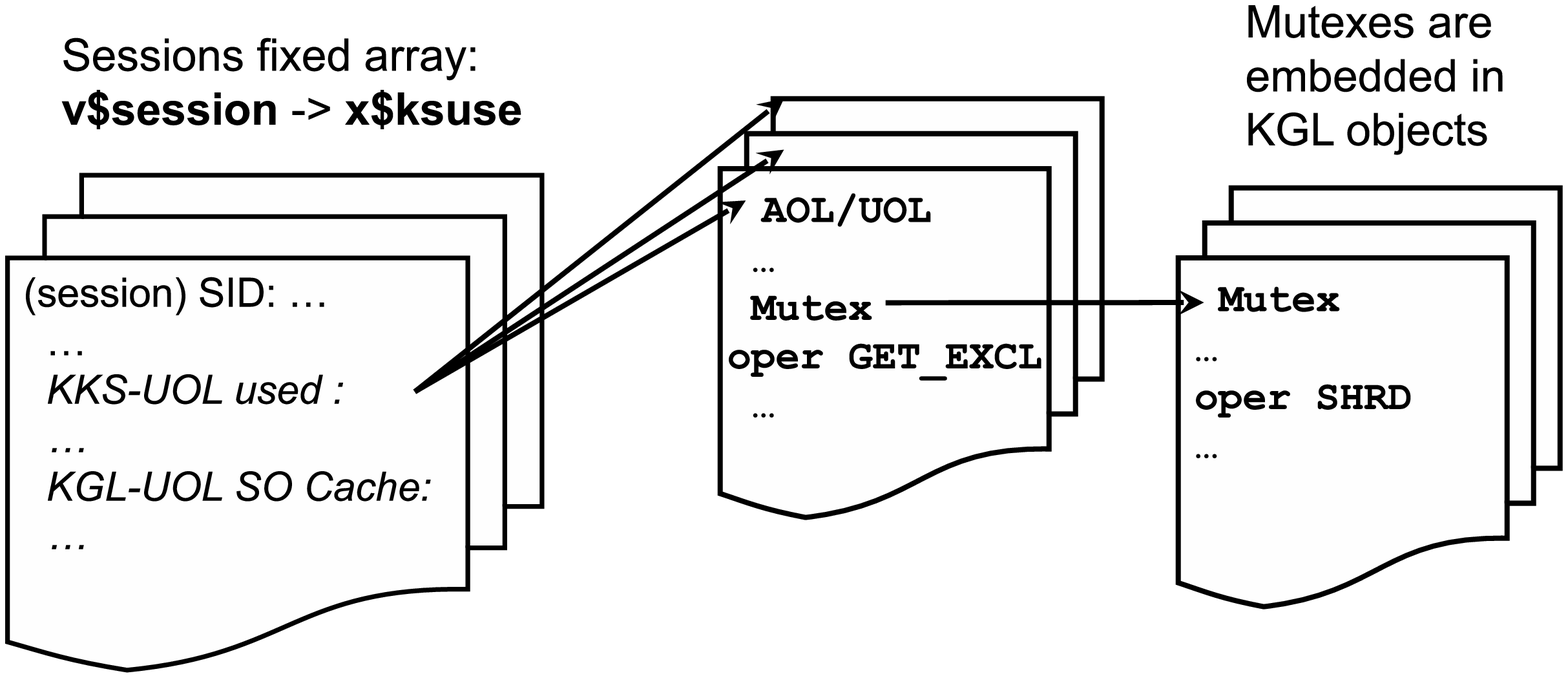}
\vspace*{-2mm}
\caption{Mutex AOL}\label{fig:maol}
\end{center}
\end{figure}
\vspace*{-6mm}

 The AOL contains information about mutex operation in progress. To operate on mutex, session first creates AOL, fills it with data about mutex and desired operation, and calls one of mutex acquisition routines.  Each session has an array of references to the AOLs it is using.  Fig. \ref{fig:maol} illustrates this. AOLs are also used during mutex recovery if session crashes.

\paragraph{Mutex  modes and states}

Mutex can be held in three modes:
\begin{itemize}
\item    {\em ''{\bf S}hared''} (SHRD in traces) mode allows mutex be holding by several sessions simultaneously. It allows read (execute) access to the structure protected by mutex. In shared mode the lower 4 bytes of mutex value represent the number of sessions holding the mutex. Upper bytes are zero.
\item    {\em ''e{\bf X}clusive'} (EXCL) mode is incompatible with all other modes. Only one session can hold the mutex in exclusive mode. It allows session exclusively access the structure protected by mutex. In {\bf X} mode upper bytes of mutex value are equal to holder SID. Lower bytes are zero.
\item {\em''{\bf E}xamine''} (SHRD\_EXAM in dumps) mode indicates that mutex or its protected structure is in transition. In {\bf E} mode upper bytes of mutex value are equal to holder SID. Lower bytes represent the number of sessions simultaneously holding the mutex in {\bf S} mode. Session can acquire mutex in {\bf E} mode or upgrade it to {\bf E} mode even if other sessions are holding mutex in {\bf S} mode. No other session can change mutex at that time.
\end{itemize}

\begin{figure}[h]
\begin{center}
\includegraphics[width=\linewidth]{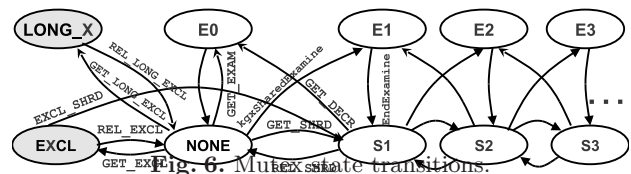}
\vspace*{-8mm}
\caption{Mutex state transitions.}\label{fig:mtrans}
\end{center}
\end{figure}

My experiments demonstrated that mutex state   transitions diagram looks like
infinite fence containing shared states 0,1,2,3,\ldots and corresponding examine states 0,1,2,3,\ldots
There are also {\em EXCL} and {\em LONG\_EXCL} states (fig. \ref{fig:mtrans}) .

Not all operations are used by each mutex type. The {\muf ''Cursor Pin''} mutex pins
the cursor in the Library Cache during parse and execution in 8-like way:
\begin{figure}[H]
\begin{center}
\includegraphics[width=\linewidth]{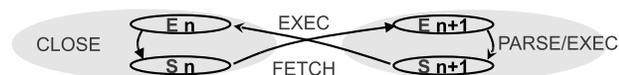}
\vspace*{-3mm}
\caption{"Cursor Pin" mutex state diagram.}\label{fig:mpin}
\end{center}
\end{figure}
\vspace*{-6mm}
Here {\bf E} an {\bf S} modes effectively act for {\muf ''Cursor Pin''} mutex as exclusive and free states.
The {\muf ''Library Cache''} mutex uses {\bf X} mode only.
This paper math is targeted on these mutexes types.

The {\muf ''hash table''} mutexes utilize both {\bf X} and {\bf S} modes. Such ''read-write'' spinlocks will be investigated in separate paper.

When  session is waiting for  mutex, it registers the wait in Oracle Wait Interface \cite{millsap}. Most frequently observed waits are  named {\bf ''cursor: pin S}, {\bf ''cursor: pin S wait on X} and {\bf ''library cache: mutex X''}. The  naming scheme is presented in table \ref{tab:wait}.
Here {\em mutex} is the name of mutex type.


\begin{table}
\centering
\small{\setlength{\extrarowheight}{1.5pt}
\begin{tabular}{|r|c|c|c|}
\hline
&get \textbf{S}&get \textbf{X,LX}&get \textbf{E}\\
\hline Held \textbf{S}&-&{\em mutex} S&?\\
\hline \textbf{ X}&{\em mutex} X&{\em mutex} X&{\em mutex} X\\
\hline \textbf{ E}&-&{\em mutex} S wait on X&{\bf {\em mutex} S}\\
\hline
\end{tabular}
}\normalsize
\caption{Mutex waits in Oracle Wait Interface}\label{tab:wait}
\end{table}

\paragraph{Experimental setup to explore mutex wait}

Unlike the latch, the details of mutex wait were not documented by Oracle. We need explore it using DTrace.
To explore the latch in \cite{latches2011}, I acquired it directly calling {\bf kslgetl} function. This is not possible for mutex. However, by changing memory I can make mutex ''busy'' artificially.
Oracle {\bf oradebug} utility allows changing of any address inside SGA:
\begin{verbatim}
SQL>oradebug poke <mutex addr> 8 0x100000001
BEFORE: [3A9371338, 3A9371340) =
                           00000000 00000000
 AFTER: [3A9371338, 3A9371340) =
                           00000001 00000001
\end{verbatim}

This looks exactly like session with SID 1 is holding the mutex in {\bf E} mode.
I wrote several scripts that simulate a busy mutex in  {\bf S}, {\bf X} and  {\bf E} modes.
In these scripts one session artificially holds  mutex for 50s.
Another session tries to acquire mutex and "statically" waits for {\bf ''cursor: pin S''} event during 49s.
DTrace allowed me explore how Oracle actually waits for mutex.

\paragraph{Original Oracle 10g mutex busy wait}

Oracle introduced the mutexes in version 10.2.0.2. Running the script  against  this version
I saw that the waiting process consumed one of my CPUs completely.
Oracle showed millions  microsecond waits that accounted for 3 seconds out of actual 49 second wait. The wait trace looks like:
\begin{alltt}
\ldots spin 255 cycles
   yield()
  spin 255 cycles
   yield()
\ldots repeated 1910893 times
\end{alltt}
The session waiting for mutex repeatedly spins 255 times polling the mutex location and then issues {\bf yield()} OS syscall. This syscall just allows other processes to run.

Oracle 10.2-11.1 counts wait time as the time spent off the CPU waiting for other processes. If the system has free CPU power, Oracle thought it was not waiting at all and mutex contention was invisible.

Therefore the old version mutex was ''classic'' spinlock without sleeps. If the mutex holding time is always small, this algorithm minimizes the elapsed time to acquire mutex. Spinning session acquires mutex immediately after its release.

Such spinlocks are vulnerable to variability of holding time. If sessions hold  mutex   for a long time, pure spinning wastes CPU. Spinning sessions can aggressively consume all the CPUs and affect the performance by priority inversion and CPU starvation.

\paragraph{Mutex wait with Patch 6904068}

If long ''cursor: pin S'' waits were consistently observed in Oracle 10.2-11.1, then  system do not have enough spare CPU for busy waiting. For such a case, Oracle provides the possibility to convert ''busy'' mutex wait into ''standard'' sleep. This enhancement was named {\em ''Patch 6904068: High CPU usage when there are ''cursor: pin S'' waits''}.  With this patch the mutex wait trace becames:
\begin{alltt}
\ldots spin 255 times
  semsys()   timeout=10 ms
\ldots repeated 4748 times
\end{alltt}
The {\bf semtimedop()} is "normal" OS sleep. The patch significantly decreases CPU consumption by spinning. Its drawback is larger elapsed time to obtain mutex. Ten milliseconds is long wait in Oracle timescale.

One can adjust  sleep time   with centisecond granularity and even set it to 0 dynamically. In such case the Oracle instance behaves exactly like without the patch. It makes sense to install the patch 6904068 in 10.2-11.1 OLTP environments proactively.

\section{IV. Mutex statistics}

Mutex statistics are the tools to diagnose  its  efficiency.  Oracle internally counts the numbers of gets and sleeps for mutex. However,  there is no fixed table containing current statistics. The {\bf x\$mutex\_sleep\_history}   shows  statistics as they were
{\em at the time of last sleep}. This is not enough.

Hopefully, Oracle provide us the {\bf x\$ksmmem} fixed table. It shows contents of any address inside SGA.  The mutex value, its {\em gets} and {\em sleeps} can be directly read out from Oracle memory. Repeatedly sampling mutex value we can estimate another key mutex statistics - {\em Utilization}. The Little's law $U=\lambda S$ allows computing the average mutex holding time $S$.

Unlike the latches \cite{latches2011}, mutex do not count its  misses and spin gets. The {\em miss ratio} $\rho$ should be estimated  from PASTA (Poisson Arrivals See Time Averages) property $\rho\approx U$.

 Oracle counts only the first mutex get, but all the secondary sleeps. Therefore, the {\em spin inefficiency} koefficient $k$ differs from experimentally observed {\em sleep ratio} $\kappa=\frac{\Delta \textrm{sleeps}}{\Delta\textrm{misses}}$.

 If  sleeps  are much longer then mutex correlation time then every {\em spin-and-sleep} cycle observe independent picture. Due to this independency each sleep has equal probability  $k\rho$ and one can estimate:
\[
\kappa=k+(k\rho)k+(k\rho)^2k+\ldots=\frac{k}{1-k\rho}
\]

 Table \ref{tab:mstats} summarizes mutex statistics and their relations. Corresponding script {\bf mutex\_statistics.sql} to measure mutex statistics is available in \cite{asn}.

\begin{table}[t]
\centering
\small{\setlength{\extrarowheight}{2.5pt}
\begin{tabular}{|p{0.33\linewidth}|p{0.23\linewidth}|p{0.3\linewidth}|}
\hline
\textbf{Description} &\textbf{Definition}&\textbf{Relations}\\
\hline  Mutex requests arrival rate&$\lambda=\frac{\Delta \textrm{gets}}{\Delta \textrm{time}}$&\\
\hline
Sleeps rate&
$\omega=\frac{\Delta \textrm{sleeps}}{\Delta \textrm{time}}$&
$\omega=\kappa\rho\lambda$\\
\hline Miss ratio (PASTA estimation)&
{$\rho=\frac{\Delta \textrm{misses}}{\Delta \textrm{gets}}$}&$\rho\approx U_X$\\
\hline Sleeps ratio&$\kappa=\frac{\Delta \textrm{sleeps}}{\Delta\textrm{misses}}$
&$\kappa=\frac{\omega}{\lambda \rho}=\frac{k}{1-k\rho}$\\
\hline Avg. holding time \hfill (Little's law)&$S=\frac{U}{\lambda}$&\\
\hline Mutex spin inefficiency&$k=\frac{\Delta \textrm{sleeps}}{\Delta \textrm{spins}}$&$k=\frac{\kappa}{1+\kappa\rho}$\\
\hline
\end{tabular}
}
\normalsize
\caption{Mutex statistics.}\label{tab:mstats}
\end{table}

 The spin time $\Delta$ can be obtained in my experiments by counting the {\em ''spin-and-yield'} cycles per second. Contemporary Oracle versions can  adjust $\Delta$ using parameter {\bf\_mutex\_spin\_count}. Therefore,  we can compute spin and yield times separately by linear regression.

Typical nocontention values for spin, yield and mutex holding time $S$ in exclusive mode on some platforms are summarized in table \ref{tab:mtimes}.

\begin{table}[h]
\centering
\small{\setlength{\extrarowheight}{1.5pt}
\begin{tabular}{|l|c|c|c|c|}
\hline
&\textbf{\scriptsize Library cache} &\textbf{\scriptsize Cursor pin}&\textbf{\scriptsize spin}&\textbf{\scriptsize yield()}\\
\hline \textbf{\scriptsize Exadata}&$0.3-5 \mu\textrm{s}$&$0.1-2\mu\textrm{s}$&$1.8\mu\textrm{s}$&$0.7\mu\textrm{s}$\\
\hline \textbf{\scriptsize Sparc T2}&$2.5-12 \mu\textrm{s}$&$3.2-11\mu\textrm{s}$&$8.7\mu\textrm{s}$
&$9.5\mu\textrm{s}$\\
\hline
\end{tabular}
}\normalsize
\caption{Average mutex spin and {\bf yield()} times.}\label{tab:mtimes}
\end{table}

Compare these microsecond times with default mutex sleep of 10 ms duration. Indeed, the mutex sleep is 10000 times longer than spin.

\section{V. ''Mean Value Analysis'' of mutex retrials}

''Mean Value Analysis'' (MVA) is an elegant approach for queuing systems invented by M. Reiser, et al. \cite{MVA}. Recent work \cite{MVAR} discussed the MVA for  retrial queues. Though not applicable directly to non-Markovian mutex, this approach can be useful for estimations.

The important point of the following approximation is replacement of fixed time mutex sleep by exponential memoryless distribution.
According to PASTA, request arriving with frequency $\lambda$ finds mutex busy with probability $\rho$ and goes to orbit (sleeps) for time $T$ with probability $k\rho$.

The waiting time consist of  spin and sleep in the orbit times.
\begin{equation}
W = W_{s} + W_{orb}
\end{equation}
The process acquires the busy mutex during repeating spins.
The total spin time is:
\begin{equation} W_s = \rho\Gamma + (k \rho) \rho \Gamma + {\left( k \rho \right)}^2 \rho \Gamma  + \ldots = \frac{\rho}{1- k \rho}\Gamma
\end{equation}

The request  retries from orbit while mutex is  busy and idle (Fig. \ref{fig:mwflow}).
\[ W_{orb} = W_b + W_i
\]

In steady state the overall busy mutex wait time is needed to serve all requests currently in system.
\[
 W_b + W_s = L_{orb}S + L_{s} S + \rho S_r
\]
Here $S_r$ is the residual mutex holding time. 
According to Little's law:
\[ L_{orb} = \lambda W_{orb},\ L_b=\lambda W_b, \ L_s=\lambda W_s, \ \rho=\lambda S.
\]
Therefore:
\begin{equation}
 W_b=\rho( S_r + W_{orb}) - (1-\rho) W_s \label{eq:mva}
\end{equation}

Flows  per second going to and from the orbit should be balanced. For exponential sleep approximation:
\[
 k \lambda\rho + k \frac{\lambda W_b}{T}=\frac{\lambda W_{orb}}{T}
\]
here $T$ is the average time to sleep.  One can substitute, in spirit of MVA, the (\ref{eq:mva})  into this expression and estimate
 the average wait time  spent on orbit as:
\[ W_{orb} =\frac{k}{1-k\rho}\left(\rho(T + S_r) - (1-\rho)W_s\right)
\]
The overall wait time  became:
\begin{equation}
W =\frac{\rho }{1- k \rho}\left(\frac{1-k^2\rho}{1- k\ \rho}\Gamma + k \left(T + T_r\right) \right),\label{eq:wmva}
\end{equation}
where $T_r$ is the residual after-spin mutex holding time that already appeared in  (\ref{eq:nocont}).

Normally in Oracle 11.2 the {spin inefficiency} $k \ll  1$ and huge sleep time $ T \sim  10^4 \times \{\Gamma, \Delta, T_r\}$  dominates in these formulas and limits the mutex wait performance
\[ W \approx\frac{k \rho }{1- k \rho}\left(T + T_r\right). \]
In order to compare this to mutex experimental data it should be noted that, unlike the queuing theory, the Oracle Wait Interface does not treat the first spin as a part of wait\cite{millsap}. The wait time registered by OWI is 
$ W_{o} = W- \rho \Gamma$.

Oracle performance tuning  uses  an {\em ''average  wait duration''} metric from AWR report \cite{Concepts112} as a contention signature. This  is the OWI waiting time normalized by the {\em number of OWI waits}:
\[ \overline{w}_{o}  = {W_o}/{k\rho} \approx \frac{1}{1-k\rho}\left( T + T_r  \right)
\]
If this quantity significantly differs from 1cs, it may be a sign of abnormality in mutex utilization or holding time.

Of course the above estimations do not account for OS scheduling and are not applicable when number of active processes exceeds the number of CPUs.

\section{VI. 11.2.0.2.2 Mutex waits diversity}

Since April 2011 the latest Oracle versions  use completely new concept of mutex waits.

{\bf My Oracle Support} site \cite{MOS} described this  in note {\em Patch 10411618 Enhancement to add different ''Mutex'' wait schemes}. The enhancement allows one of three concurrency wait schemes and introduces 3 parameters to control the mutex waits:
\begin{description}
\item{\bf \_mutex\_wait\_scheme} --- Which wait scheme to use:
\begin{description}
\item{\bf 0} - Always YIELD.
\item{\bf 1} - Always SLEEP for {\bf \_mutex\_wait\_time}.
\item{\bf 2} - Exponential Backoff upto {\bf \_mutex\_wait\_time}.
\end{description}
\item{\bf \_mutex\_spin\_count} --- the number of times to spin. Default value is 255.
\item{\bf \_mutex\_wait\_time} --- sleep timeout depending on scheme. Default is 1.
\end{description}
The note also mentioned that this fix effectively supersedes the patch 6904068  described above.

\paragraph{The SLEEPS. Mutex wait scheme 1}

In mutex wait scheme 1 session repeatedly requests 1 ms sleeps:
\begin{verbatim}
kgxSharedExamine(...)
  yield()
  pollsys() timeout=1 ms repeated 25637 times
\end{verbatim}
The {\bf \_mutex\_wait\_time} parameter controls the sleep timeout in {\em milliseconds}.
This scheme differs from patch 6904068 by one additional {\em spin-and-yield} cycle at the beginning and smaller timeout. 

\begin{figure}[h]
\begin{center}
\includegraphics[width=\linewidth]{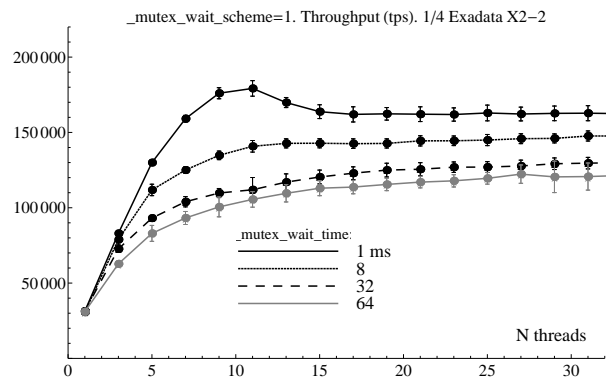}
\end{center}
\vspace{-5mm}
\caption{Mutex wait scheme 1 throughput.}\label{fig:wait1}
\end{figure}

Performance of this scheme is sensitive to {\bf \_mutex\_wait\_time} tuning. fig. \ref{fig:wait1} demonstrates that at moderate concurrency the short mutex sleeps  performs better and results in bigger throughputs. However, such millisecond sleep will be rounded to centisecond on  platforms like Solaris, Windows and latest HP-UX. This is because most OS can not sleep for very short times.

MVA estimation for mutex wait scheme 1 results in:
\[
 W_1 \approx\frac{k \rho }{1- k \rho}\left(k T + T_r\right).
\]

You see that additional spin at the beginning effectively reduces wait time $T$ multiplying it by $k\ll 1$. This increases the performance.

\paragraph{Default ''Exponential Backoff'' scheme 2}

Oracle uses the scheme 2 by default. This scheme is named ''Exponential backoff'' in  documentation. Unlike the previous versions, contemporary mutex wait do not consumes CPU. Surprisingly, DTrace shows that there is no exponential behavior by default. Session repeatedly sleeps with 1 cs duration:
\begin{verbatim}
 yield() call repeated 2 times
 semsys() timeout=10 ms repeated 4237 times
\end{verbatim}
To reveal exponential backoff one need to increase the {\bf \_mutex\_wait\_time} parameter.
\begin{verbatim}
SQL> alter system set "_mutex_wait_time"=30;
...
 yield() call repeated 2 times
 semsys() timeout=10 ms repeated 2 times
 semsys() timeout=30 ms repeated 2 times
 semsys() timeout=80 ms
 semsys() timeout=70 ms
 semsys() timeout=160 ms
 semsys() timeout=150 ms
 semsys() timeout=300 ms repeated 159 times\end{verbatim}
This closely resembles the Oracle 8i latch acquisition algorithm \cite{latches2011,asn}.
In scheme 2 the {\bf \_mutex\_wait\_time} controls  maximum wait time in {\em centiseconds}. Due to exponentiality the mutex wait scheme 2 is insensitive to  its value. Indeed, only sleep after the fifth unsuccessful spin is affected by this parameter \cite{asn}.

\begin{figure}[t]
\begin{center}
\vbox{
\includegraphics[width=0.49\linewidth]{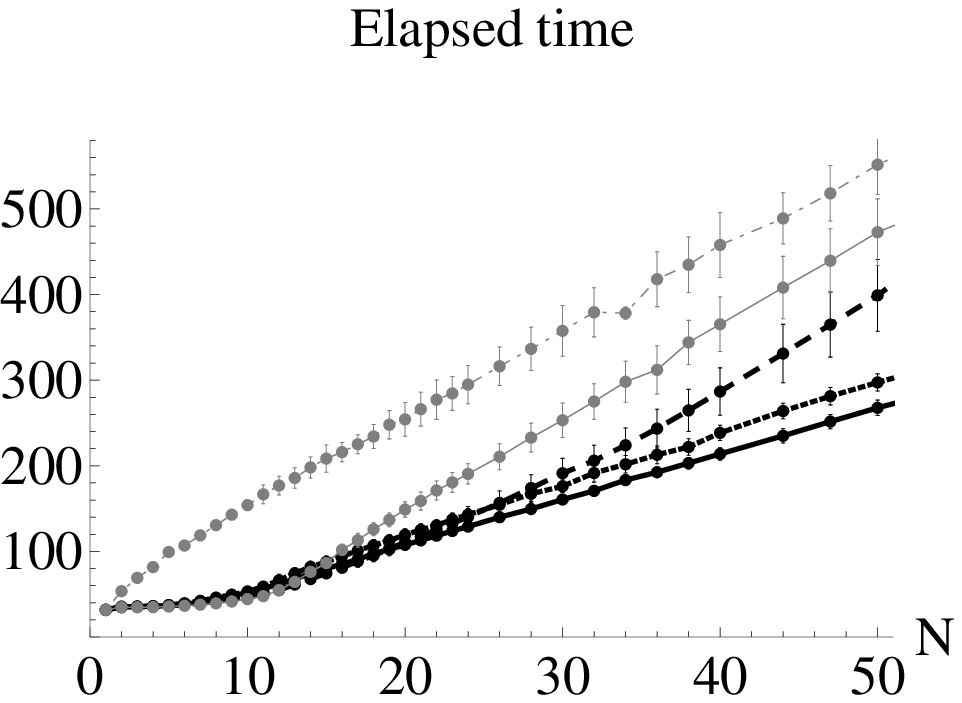}
\includegraphics[width=0.49\linewidth]{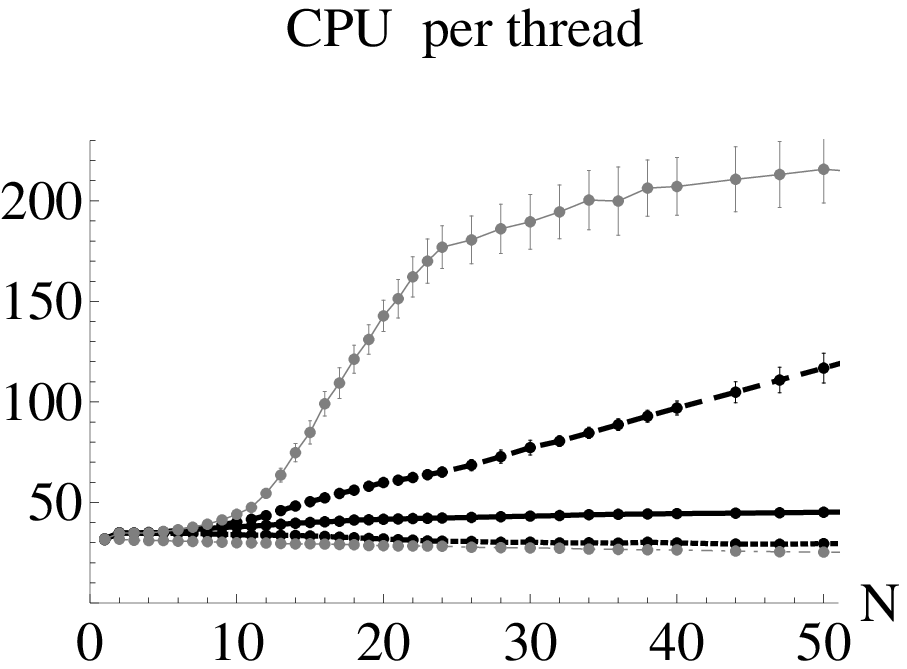}}
\vspace{3mm}
\end{center}
\includegraphics[width=\linewidth]{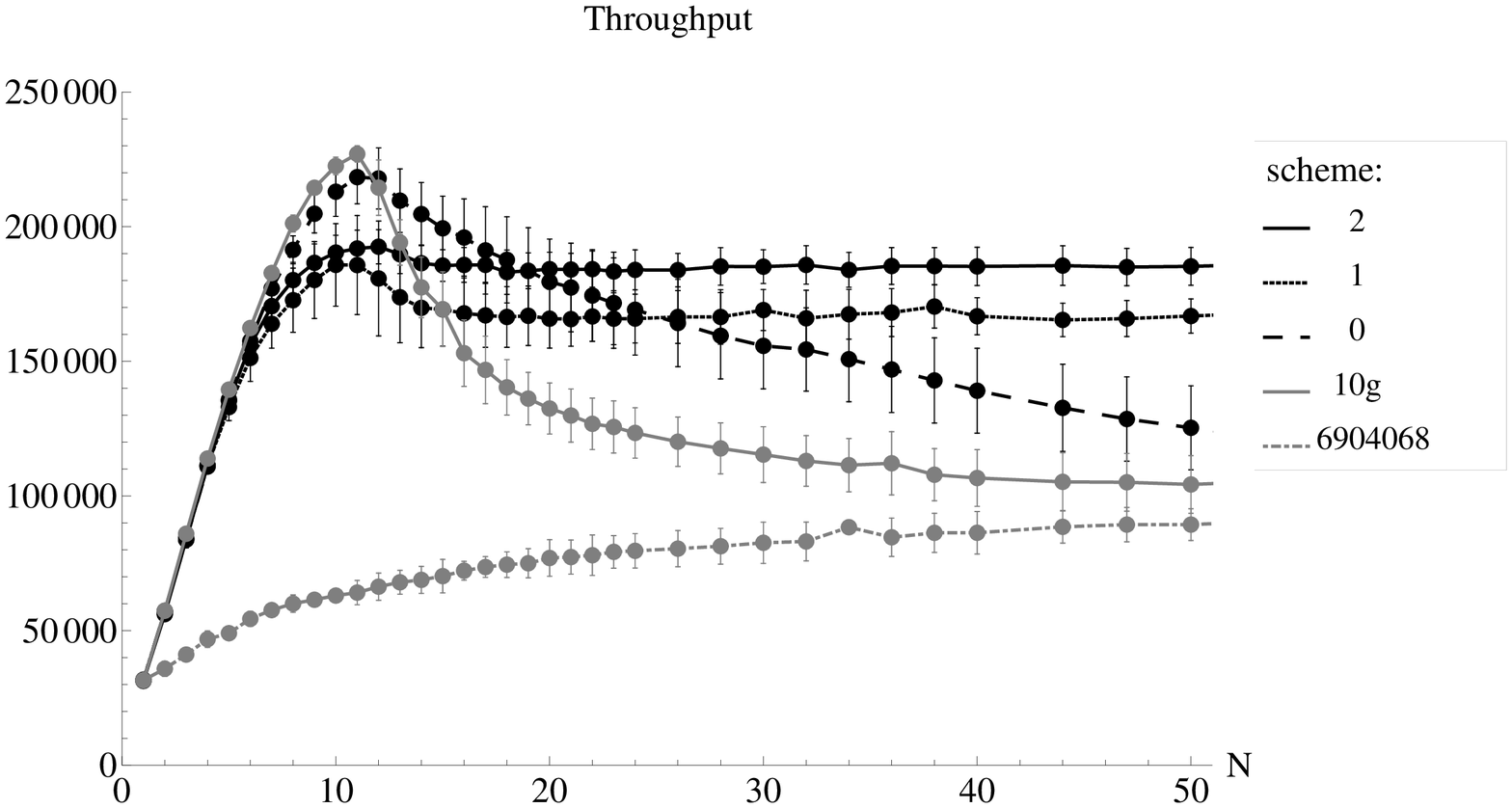}
\vspace{-5mm}
\caption{Mutex wait schemes performance.}\label{fig:wait5}
\end{figure}

Default mutex  scheme 2 wait differs from patch 6904068 by two {\bf yield()} syscalls at the beginning.
These two {\em spin-and-yields} change the mutex wait performance drastically (fig. \ref{fig:wait5}). They   effectively multiply centisecond wait time $T$ by $k^2$:
\[ W_2 \approx\frac{k\rho }{1- k \rho}(k^2 T + T_r).
\]
\vspace{5mm}

\paragraph{Classic YIELDS. Mutex wait scheme 0}

The {\bf \_mutex\_wait\_scheme} 0 consists  mostly of repeating {\em spin-and-yield} cycles.
\begin{verbatim}
 yield() call repeated 99 times
 pollsys() timeout=1 ms
 yield() call repeated 99 times
 pollsys() timeout=1 ms
... \end{verbatim}
  It differs from aggressive mutex waits used in previous Oracle versions by 1ms sleep after each 99 yields. This sleep significantly reduces CPU consumption and increases robustness. Unfortunately previous MVA style analysis is not applicable for this wait scheme.

The scheme 0 is very flexible \cite{asn}.  The sleep duration and yield frequency are tunable by {\bf \_wait\_yield\_sleep\_time\_msecs} \ldots parameters.
   One can also specify different wait modes for standard and high priority processes.  This allows almost any combination of yield and sleeps including 10g and patch 6904068 behaviors.

\paragraph{Comparison of mutex wait schemes}

Fig. \ref{fig:wait5} compares performance of  {\muf ''Library Cache''} mutex contention testcase on Exadata platform for all 3  wait schemes and the {\em patch 6904068} and {\em 10g} mutex algorithms as well.  

The figures demonstrate that:
\begin{description}
\item{\bf Default scheme 2}  is well balanced in all concurrency regions.
\item{\bf Wait scheme 1} should be used when the system is constrained by CPU.
\item{\bf Wait scheme 0} has the throughput close to 10g in medium concurrency region and may be recommended
in case of plethora of free CPU.
\item{\bf 10g} mutex algorithm had the fastest performance in medium concurrency workloads.
However, its throughput fell down when number of contending threads exceeds number of CPU cores. CPU consumption increased rapidly beyond this point.
This excessive CPU consumption starves processors and impacts other database workloads.

\item {\bf Patch 6904068} results in very low CPU consumption, but the largest elapsed time and the worst throughput.
\end{description}

\section{IV. Mutex Contention}
Mutex contention occurs when the mutex is requested by several sessions at the same time. Diagnosing mutex contention we always should remember Little's law
\[ U=\lambda S \]
Therefore, the contention can be consequence of either:

 Long mutex holding time $S$ due to, for example,  high SQL version count, bugs causing long mutex holding time or CPU starvation and preemption issues.

Or  it may be due to high mutex exclusive Utilization.   Mutexes may be overutilized by too high SQL and PL/SQL execution rate or
 bugs causing excessive requests.

Mutex statistics help to diagnose what actually happens.

Latest Oracle versions include many fixes for  mutex related bugs and allow flexible  wait schemes,
adjustment of  spin count and  cloning of hot library cache objects. My blog \cite{asn} continuously  discusses related enhancements.

 Traditionally tuning of mutex performance problems was focused  on changing the application  and reducing the mutex demand. To achieve this one need to tune the SQL operators, change the physical schema, raise the bug with Oracle Support, etc\ldots  \cite{lewis,adams,millsap,lp}.

However, such tuning may be too expensive and even require complete application rewrite. This article discusses  one not widely used tuning possibility - changing of mutex spin count. This was commonly treated as an old style tuning, which should be avoided by any means. The public opinion is that increasing of spin count leads to waste of CPU. However, nowadays the CPU power is cheap. We may already have enough free  resources. It makes sense to know  when the spin count tuning may be beneficial.

\paragraph{Mutex spin count tuning}

Long mutex holding time may cause the mutex contention.
Default {\bf \_mutex\_spin\_count} =255 may be too small.
Longer spinning may alleviate this.
If the mutex holding time distribution has exponential tail:

\begin{displaymath}
\left.
\begin{array}{l}
\mathsf{Q}(t) \sim C \exp(-t/\tau)\\
k \sim C \exp(-t/\tau) \\
\Gamma\sim S_r - C \tau \exp(-t/\tau)\\
\end{array} \right.
\end{displaymath}
It is easy to see that if "sleep ratio" is small enough ($k \ll 1$) then

{\em Doubling the spin count will square the ''sleep ratio'' and will only add part of order of  $k$    to spin CPU consumption.}

\begin{figure}[t]
\begin{center}
\vbox{
\includegraphics[width=0.46\linewidth]{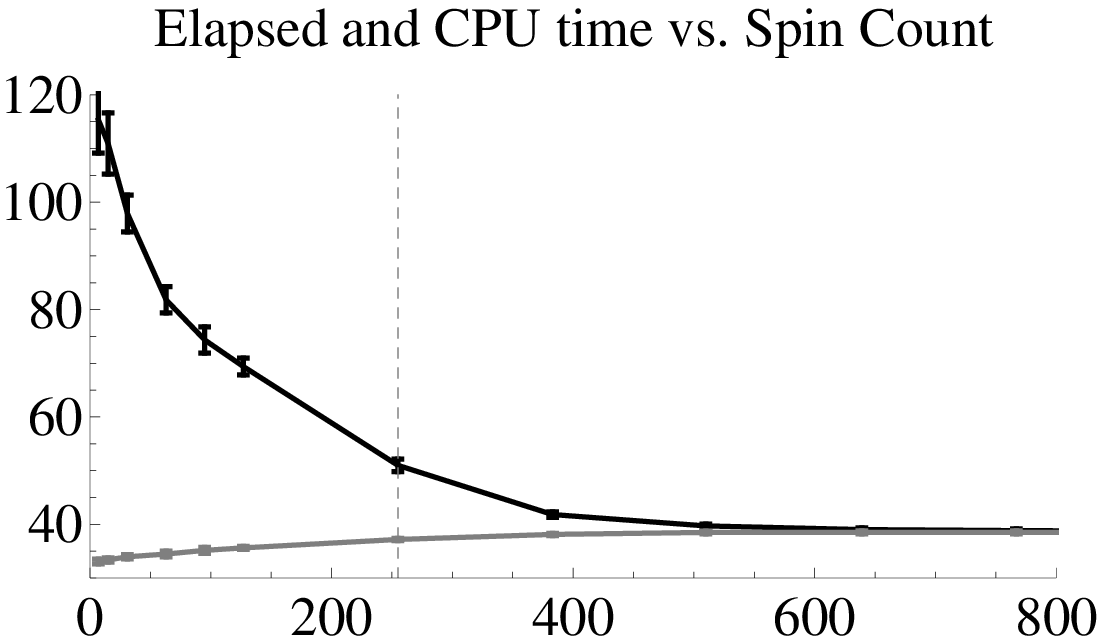}
\includegraphics[width=0.52\linewidth]{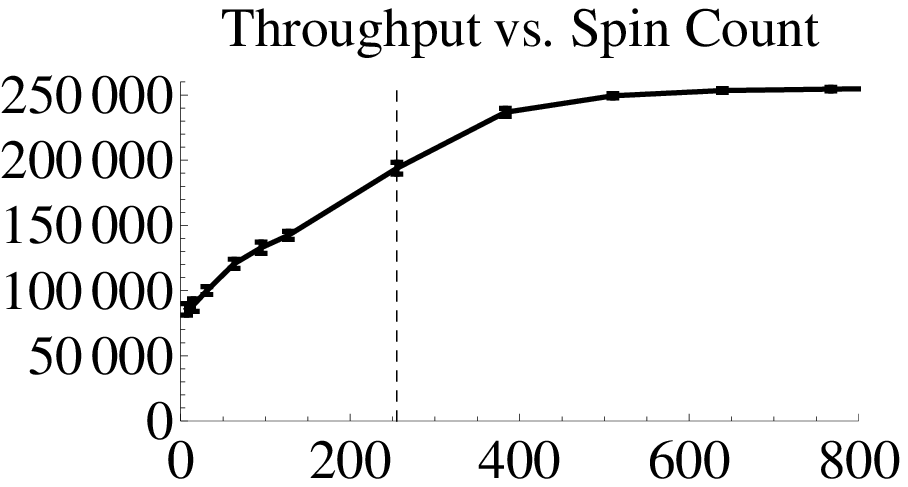}}
\vspace{3mm}
\vbox{
\includegraphics[width=0.47\linewidth]{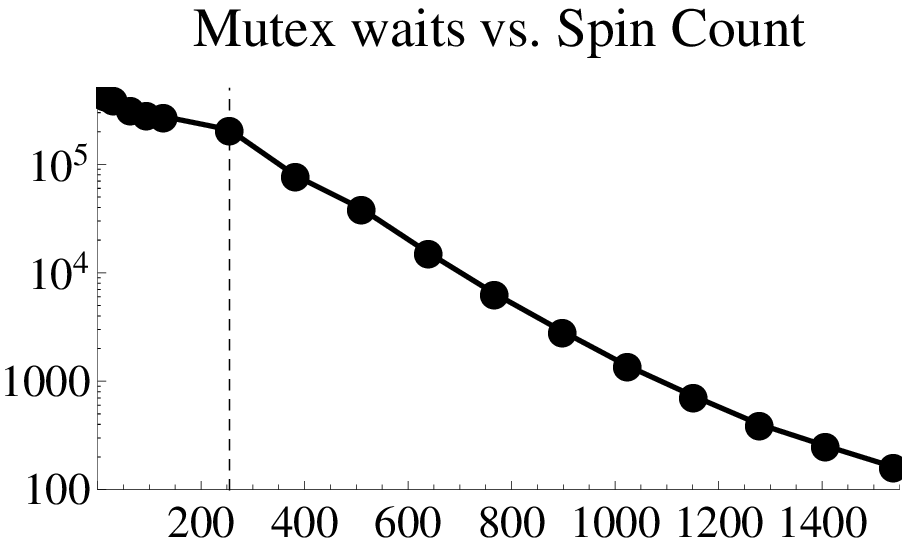}
\includegraphics[width=0.47\linewidth]{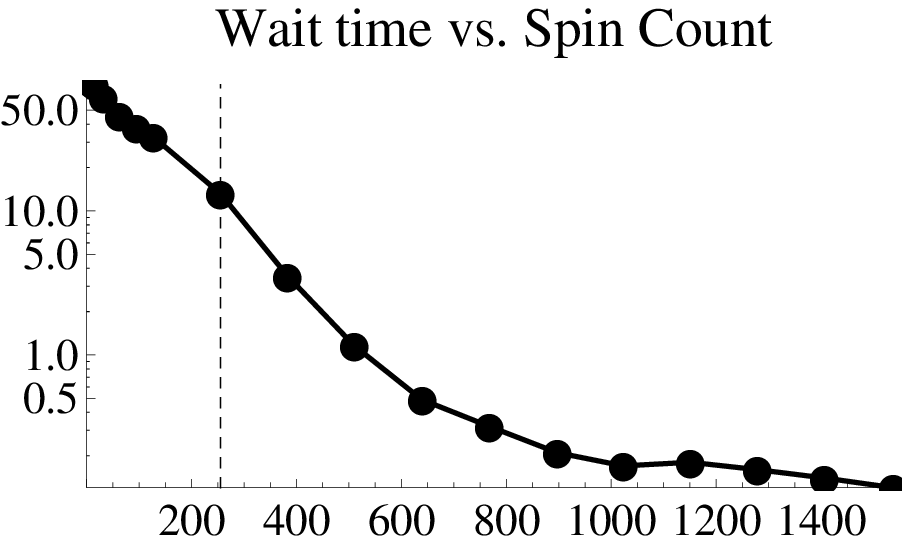}}
\end{center}
\vspace{-5mm}
\caption{Spin count tuning.}\label{fig:spincount}
\end{figure}

In other words, if the spin is already efficient, it is worth to increase the spin count. Fig. \ref{fig:spincount} demonstrates effect of  spin count adjustment for  the {\muf ''Library Cache''} mutex contention testcase.


The spin count tuning is very effective.
Elapsed time fell rapidly  while CPU increased smoothly.
The number of mutex waits demonstrates almost linear behavior in logscale. This confirms the scaling rule.

\section{Conclusions}

This work investigated the possibilities to diagnose and tune mutexes, retrial Oracle spinlocks.
Using DTrace, it explored how the mutex works, its spin-waiting schemes, corresponding parameters and statistics. The mathematical model was developed to  predict  the effect of mutex tuning.

The results are important for  performance tuning of highly loaded Oracle OLTP databases.

\section{Acknowledgements}
Thanks to Professor S.V.\ Klimenko for kindly inviting me to MEDIAS 2012 conference.

Thanks to RDTEX CEO I.G.\ Kunitsky for financial support.
Thanks to RDTEX Technical Support Centre Director S.P.\ Misiura for years of encouragement and support of my investigations.

Thanks to my colleagues for discussions and all our customers for participating in the mutex troubleshooting.

\bibliographystyle{IEEEbib}

\section*{About the author}

Andrey Nikolaev is an expert at RDTEX First Line Oracle Support Center, Moscow. His contact email is \texttt{Andrey.Nikolaev@rdtex.ru}.

\end{document}